\documentclass[journal=nalefd,manuscript=article]{achemso}
\usepackage[version=3]{mhchem} 
\usepackage{graphicx}
\usepackage{color}
\usepackage{ulem}
\usepackage{braket}

\author{Benjamin\ W.\ Heinrich}
\email{bheinrich@physik.fu-berlin.de}
\author{Lukas\ Braun}
\altaffiliation{present address: Fritz-Haber-Institut der Max-Planck-Gesellschaft, 14195 Berlin, Germany}
\author{Jose\ I.\ Pascual}
\affiliation{Fachbereich Physik, Freie Universit\"at Berlin, Arnimallee 14, 14195 Berlin, Germany}
\alsoaffiliation{CIC nanoGUNE, 20018 Donostia-San Sebasti\'an, Spain}
\alsoaffiliation{Ikerbasque, Basque Foundation for Science, 48011 Bilbao, Spain}
\author{Katharina\ J.\ Franke}
\affiliation{Fachbereich Physik, Freie Universit\"at Berlin, Arnimallee 14, 14195 Berlin, Germany}

\date{\today}

\title{Tuning the magnetic anisotropy of single molecules}

\abbreviations{STM, BCS, Fe-OEP-Cl, Fe-OEP, ISTS}

\begin{document}

\begin{abstract}
{The magnetism of single atoms and molecules is governed by the atomic scale environment. In general, the reduced symmetry of the surrounding splits the $d$ states and aligns the magnetic moment along certain favorable directions. Here, we show that we can reversibly modify the magnetocrystalline anisotropy by manipulating the environment of single iron(II) porphyrin molecules adsorbed on Pb(111) with the tip of a scanning tunneling microscope.
When we decrease the tip--molecule distance, we first observe a small increase followed by an exponential decrease of the axial anisotropy on the  molecules. This is in contrast to the monotonous increase observed earlier for the same molecule with an additional axial Cl ligand~\cite{heinrich13NP}.  
We ascribe the changes in the anisotropy of both species to a deformation of the molecules in the presence of the attractive force of the tip, which leads to a change in the $d$ level alignment. 
These experiments demonstrate the feasibility of a precise tuning of the magnetic anisotropy of an individual molecule by mechanical control.}
\end{abstract}

{KEYWORDS: {magnetocrystalline anisotropy, spin excitation, strain-induced tuning, porphyrins, manipulation, scanning tunneling microscopy}}
\\

An important goal in device technology is the construction of nanomagnets and addressability of their magnetic states~\cite{gatteschi06,affronte08}. 
This requires the understanding of how the atomic-scale environment influences the spin state and magnetocrystalline anisotropy of single atoms, the building blocks of nanomagnets~\cite{gambardella03,miyamachi13,bryant13,rau14}.
The  spin state of a magnetic atom is governed by the  competition of two energy scales: the  splitting  of the $d$ orbitals by the reduced symmetry in their crystal field, which tends to lower the total spin, and the tendency of electron spins to avoid spin pairing and align according to Hund's rules.
In a metal-organic molecule, the $d$ level splitting is adjusted by the organic ligand around the paramagnetic metal atom.
The magnetocrystalline anisotropy is a consequence of spin-orbit coupling via the admixture of low-lying excited states as a second order perturbation~\cite{wang93}.
Recent works have shown that the careful choice of the atomic surrounding on surfaces, such as adsorption sites~\cite{gambardella03,honolka12,miyamachi13,bryant13,rau14}, or an organic ligand~\cite{tsukahara09,gambardella09,heinrich13NL} leads to sizable magnetocrystalline anisotropy, up to some tens of meV. 

Tunable control of the magnetocrystalline anisotropy is difficult to achieve, because it demands a handle to continuously shift the energy levels~\cite{bhandary11}. 
So far, this has only been achieved in mechanically controlled break junctions~\cite{parks10}, which lack the precise control of the molecular geometry and, hence, of the type of deformation induced.
One way to tune properties of nanoobjects on surfaces with a high degree of control consists in approaching the tip of an STM to the object and modifying the local magnetic, electronic or mechanical properties. 
This has been used to intentionally shift surface~\cite{limot03}, or molecular states~\cite{neel08} in energy, alter the Kondo screening~\cite{choi12}, or the \mbox{(anti)}ferromagnetic coupling strength~\cite{bork11}, and change the spin state mixing of individual nanomagnets~\cite{yan14}. 
 
Here, we develop a strategy to precisely tune the magnetocrystalline anisotropy in metal-organic complexes. We study and manipulate iron(II) porphyrin molecules adsorbed on Pb(111) by means of scanning tunneling microscopy and inelastic scanning tunneling spectroscopy (ISTS) and compare the results to measurements on the same molecule with an additional axial Cl ligand.
We use inelastic spin excitations to monitor changes in the magnetic anisotropy. To be able to resolve even small variations, a very good energy resolution is required. Thus, we employ a superconducting tip and substrate to probe the magnetic excitations in the single molecules. We then use the force exerted by the tip of our microscope to affect the molecules with picometer precision. We gain a handle on the ligand field splitting and, hence, the magnetocrystalline anisotropy.
These findings pave the way for a flexible atomic-scale control of spins in nanomagnets.

\begin{figure}[t]
  \includegraphics[width=0.65\textwidth,clip=]{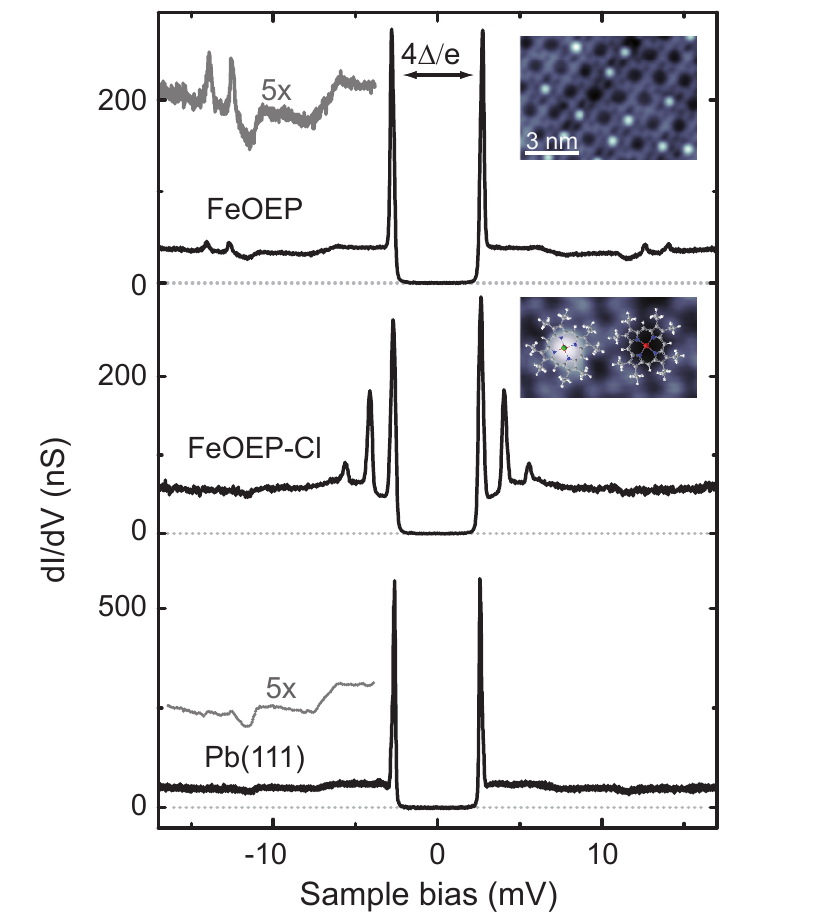}
  \caption{ISTS on Fe-OEP on Pb(111). The $dI/dV$ spectra are acquired above the center of an Fe-OEP, and an Fe-OEP-Cl molecule, respectively, and on pristine Pb(111), using conventional lock-in technique at a frequency of $912\,{\rm Hz}$ with an \textit{ac} bias modulation of $50\,\mu {\rm V_{rms}}$. The insets show a topographic image of a molecular island on Pb(111) ($V=130\,{\rm mV}$, $I=60\,{\rm pA}$) and a high resolution zoom on two molecules. Molecules with an axial Cl ligand appear with a protrusion in the center (Fe-OEP-Cl, left in the zoom), molecules without Cl ligand exhibit a depression in the center (Fe-OEP, right). 
}
\label{fig:topo}
\end{figure}

We carry out our experiments in a SPECS JT-STM operating at a base temperature of $1.2\,{\rm K}$. 
The Pb(111) surface was cleaned by  repeated sputter/anneal cycles until a clean, superconducting surface was obtained~\cite{note1}.   
To increase the energy resolution beyond the Fermi-Dirac-limit of normal metal tips, we covered the bulk \mbox{W-tip} with superconducting Pb by controlled indentation into the clean surface, until the tip showed a bulk-like superconducting order parameter $\Delta$. 
Iron-octaethylporphyrin-chloride (Fe-OEP-Cl, structure model as inset in Figure~\ref{fig:topo}) was sublimated from a crucible at $490\,{\rm K}$ onto the clean Pb(111) surface held at $120\,{\rm K}$. After annealing to $240\,{\rm K}$, the porphyrin molecules self-assemble in ordered monolayer islands of quasi-hexagonal structure, with the ethyl-groups clearly visible in the STM images (see insets in Figure~\ref{fig:topo}). 
About two thirds of the molecules exhibit a depression in the center. They have lost the axial Cl ligand upon adsorption (in the following referred to as Fe-OEP).  The second type of molecules shows a bright protrusion in the center, which we identify as the Cl ligand. By means of a voltage pulse, the ligand can controllably be removed (compare to Ref.~\cite{gopakumar12,stroz12}), and Fe-OEP-Cl is transformed into Fe-OEP. 


To characterize the spin state of Fe-OEP, we recorded spectra of the differential conductance ($dI/dV$) with the tip placed over the center of the molecule (Figure~\ref{fig:topo} top). 
The spectrum exhibits a gap region around the Fermi level, enclosed by sharp resonances of quasi-particle excitations at $eV=\pm 2\Delta$ as also observed for the pristine Pb(111) surface (Figure~\ref{fig:topo} bottom)~\cite{note3}. This characterizes the unperturbed superconductor-superconductor tunneling junction. No inner-gap bound states, which are ascribed to local screening of the spin through the superconductor's quasi-particles~\cite{shiba68,franke11}, are observed on the paramagnetic molecule. This shows the absence of any sizable interaction with the substrate, probably due to the eight ethyl groups acting as spacers. 
Instead, we found two pairs of new resonances appearing symmetric to zero bias at $\pm 12.6$ and $\pm 14.0$~mV, outside the superconducting gap. 
On Fe-OEP-Cl, we observed similar peak shapes (see Figure~\ref{fig:topo} middle) at lower energies~\cite{heinrich13NP}. 
The resonances are signatures of inelastic excitations in a superconductor-superconductor junction~\cite{heinrich13NP}. The opening of the inelastic tunneling channel at the threshold energy $e|V|=2\Delta + \varepsilon$ is reflected in the $dI/dV$ spectra by BCS-like resonances due to the structure of the superconducting  density of states  of tip and sample. 

\begin{figure}[t]
  \includegraphics[width=0.65\textwidth,clip=]{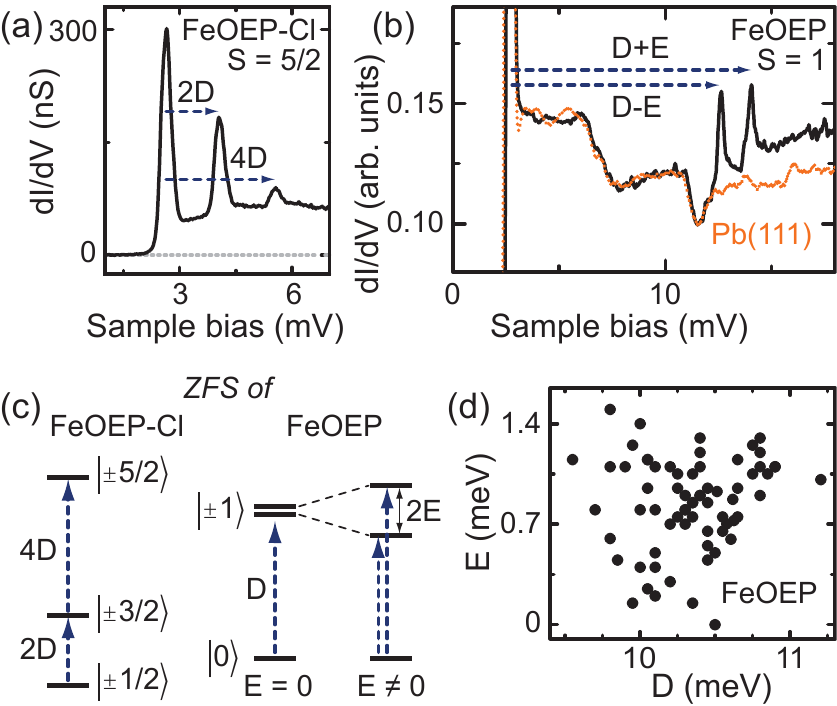}
  \caption{Inelastic excitations. (a) High-resolution ISTS spectrum of Fe-OEP-Cl ($25\,\mu {\rm V_{rms}}$). The two excitations with an energy of $1.4$ and $2.8\,{\rm meV}$ identify the $S=5/2$ state with a magnetocrystalline anisotropy parameter $D=0.7\,{\rm meV}$~\cite{heinrich13NP}. 
(b) The zoom on the ISTS spectrum of Fe-OEP unveils two inelastic excitations of equal intensity at $12.6$ and $14.0\,{\rm mV}$ ($50\,\mu {\rm V_{rms}}$). As guide for the eye, the spectrum of pristine Pb(111) is superimposed as dotted line. 
(c) Scheme of the zero field splitting of Fe-OEP and Fe-OEP-Cl. For Fe-OEP we detect, additionally to the main anisotropy axis, an in-plan distortion (rhombicity), which gives rise to the parameter $E$. 
(d) Distributions of $E$ and $D$ as measured for 71 different Fe-OEP molecules (measured at $4.5\,{\rm K}$).  
}
\label{ZFS}
\end{figure}

Measurements on Fe-OEP-Cl in a magnetic field of up to $3$~T perpendicular to the sample surface (Supporting Information, Figure~S1) show that the origin of these peaks is magnetic. 
The observed Zeeman shift of the low energy excitation and the appearance of a new transition around the Fermi level  reveals a half integer spin state of either $S=3/2$ or $5/2$ with an easy-plane anisotropy.
Current-dependent changes of the excitation intensities allowed us to identify the excitation scheme of the molecule [illustrated in Figure~\ref{ZFS}(c)]~\cite{heinrich13NP}. 
The lower-energy excitation corresponds to transitions between the ground state ($\ket{M_S}=\ket{\pm 1/2}$) and the first excited state ($\ket{\pm 3/2}$). The second excitation then proved the $S=5/2$ state, because it is due to transitions between $\ket{\pm 3/2}$ and $\ket{\pm 5/2}$, the first and second excited state, respectively. 
The deduced easy-plane anisotropy can be described in terms of the phenomenological Spin--Hamiltonian at zero magnetic field: $\mathcal{H} = DS^2_z + E(S^2_x-S^2_y)$ (here $S_x$, $S_y$, and $S_z$ are the spin operators of the three Cartesian axis)~\cite{gatteschi06}. The axial anisotropy parameter $D=0.7\,{\rm meV}$ is close to the bulk value~\cite{nishio01}. We did not detect any additional in-plane distortion ($E=0$).

The removal of the Cl ligand causes a drastic change in the excitation spectrum [Figure~\ref{ZFS}(b)].  
The two excitations at $12.6$ and $14.0\,{\rm mV}$ are of equal intensity independent of the tunneling current~\cite{note2}.  This spectrum agrees with a spin $S=1$, which is expected for an Fe center reduced from oxidation state III to II. The intermediate spin state is typical for ${\rm Fe}^{2+}$ in the square-planar ligand field of a porphyrin ligand~\cite{handbook,bernien07,wende07}.  
In the $S=1$ state, easy-plane anisotropy  ($D>0$) with an additional in-plane distortion ($E\neq 0$) gives rise to two possible excitations with equal transition probabilities [sketch in Figure~\ref{ZFS}(c)]. From the spectrum we deduce the anisotropy parameters $D=10.6\,{\rm meV}$ and $E=0.7\,{\rm meV}$. The $D$ value is similar to those measured for other Fe$^{2+}$ porphyrins in the bulk phase~\cite{boyd79,barrac70}.
However, we find variations in $E$ and $D$ around these values [Figure~\ref{ZFS}(d)] for different Fe-OEP molecules on the surface without any obvious correlation between the two anisotropy parameters (note that no such variations are observed for Fe-OEP-Cl). 
The fluctuation in $E$ and $D$ values are probably due to slight variations in the adsorption site within the self-assembled islands.

To gain an intuitive picture of the origin of the magnetocrystalline anisotropy of the two molecules, we consider the different ligand field splitting and $d$ level occupation of the Fe ion in the two complexes (Figure~\ref{fig:LFS}).
The ligand field experienced by the Fe$^{3+}$ in \mbox{Fe-OEP-Cl} is of square-pyramidal nature. The axial Cl ligand, which is a weak ligand compared to the pyrole nitrogen, lifts the Fe$^{3+}$ out of the macrocycle plane~\cite{handbook}. 
This leads to a small ligand field splitting and yields a high-spin $S=5/2$ ground state with singly occupied $d$ levels. In this case, the spin-pairing energy is larger than the energy level difference. 
The magnetocrystalline anisotropy results from the admixture of electronically excited states of reduced total spin [{\it e.g.} E$_1$ and E$_2$ in Figure~\ref{fig:LFS}(a)]. Furthermore, the lifting of the Fe center out of the pyrole plane leads to a relatively large Fe--surface distance, which makes the Fe$^{3+}$ insensitive to variations of the adsorption site.

In the case of Fe-OEP, the Fe$^{2+}$ lies in the square-planar ligand field of the porphyrin macrocycle. The in-plane position of the Fe$^{2+}$ yields shorter Fe--N distances and a larger ligand field splitting, in particular the energy of the $d_{x^2-y^2}$ level is strongly increased compared to Fe-OEP-Cl. 
This results in the intermediate spin ($S=1$) ground state E$_0$ as depicted in Figure~\ref{fig:LFS}(b).  Since the splitting between the $d_\pi$ and $d_{z^2}$ is small in Fe(II) porphyrins~\cite{note9}, 
there is a strong admixture of E$_1$ and E$_2$ to the ground state and hence a large magnetocrystalline anisotropy.
The admixture of these lowest lying states yields contributions to the orbital moments $\Lambda_{xx}$ and $\Lambda_{yy}$ in $x$ and $y$ direction, respectively, but not to $\Lambda_{zz}$, the orbital moment in $z$ direction~\cite{dai08}. Therefore, the resulting anisotropy is easy-plane, {\it i.e.}, $D>0$, as we can write \mbox{$D=-\frac{\lambda^2}{2} (2\Lambda_{zz}-\Lambda_{xx}-\Lambda_{yy})$}~\cite{dai08}, with $\lambda$ being the spin-orbit coupling constant. 
At the same time, the shorter Fe--surface distance compared to Fe-OEP-Cl results in a higher sensitivity to variations in the adsorption site. 
The atomic environment underneath the Fe$^{2+}$ ion mainly affects the $d_{z^2}$ and $d_\pi$ levels, because they extend toward the surface, resulting in slight variations of $\Lambda_{xx}$ and $\Lambda_{yy}$, and, therefore, $D$ and $E$ [note: {$E=-\frac{\lambda^2}{2} (\Lambda_{xx}-\Lambda_{yy})$]. The interaction with the surface also explains the reduction of the $D_{4h}$ symmetry evidenced by the non-zero $E$ parameter.

\begin{figure}[t]
  \includegraphics[width=0.65\textwidth,clip=]{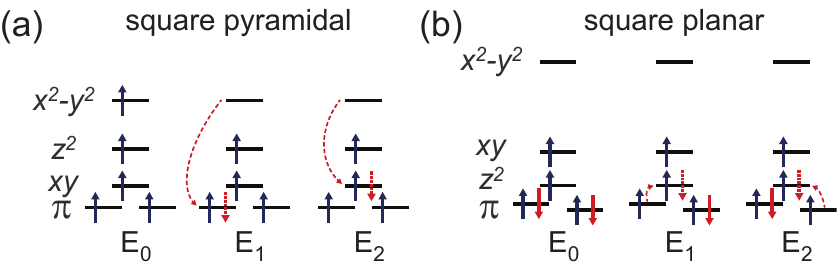}
  \caption{Sketch of the electronic ground (E$_{0}$), first (E$_{1}$), and second (E$_{2}$) excited state in the case of: (a) high-spin ($S=5/2$) Fe$^{3+}$ in a square-pyramidal ligand field (as in Fe-OEP-Cl); (b) intermediate spin ($S=1$) Fe$^{2+}$ in a square-planar ligand field~\cite{huheey03} (as in Fe-OEP). Note that in (a) E$_1$ (degenerate) and E$_2$ are of reduced total spin compared to E$_0$, while in (b) the spin is conserved. Energies not to scale.
}
\label{fig:LFS}
\end{figure}

The above shown sensitivity of the zero field splitting to small variations in the environment provides access to a controlled tuning. 
With this intention, we approach the tip of the STM to the center of an Fe-OEP molecule.
The presence of the tip alters the ligand field, while we simultaneously record ISTS curves at varying tip--sample distances [see Figure~\ref{fig:dz}(a)]. 
First, both excitation peaks shift to higher energies and reach a maximum at $\Delta z\approx-200\,{\rm pm}$. With a further reduction of the distance, both peaks shift to energies lower than the initial values until, at $-330\,{\rm pm}$, the junction becomes instable. Figure~\ref{fig:dz}(b) shows the extracted values of $D$ {\it vs}. $\Delta z$~\cite{note8}. From $\Delta z = 0$ to $-200\,{\rm pm}$ the axial anisotropy $D$ slightly increases, before it rapidly decreases for shorter distances.  
This variation of $D$ is qualitatively different from the variation observed earlier on Fe-OEP-Cl~\cite{heinrich13NP}, where $D$ exponentially increased with decreasing tip--sample distance.

These variations in the zero field splitting of the two species can be understood considering the changes in geometry induced by the proximity of the STM tip.  
In the relaxed adsorption state of Fe-OEP (in absence of the STM tip), the Fe ion is expected to be attracted toward the surface, thus lying slightly below the porphyrin plane. 
The tip potential then exerts an opposed attractive force on the Fe atom, which pulls it toward the opposite side of the macrocycle due to a surface trans effect~\cite{flechtner07}. 
Passing through the molecular plane causes first a small decrease in Fe--N bond length, followed by a subsequent Fe--N bond elongation.  
A larger Fe--N bond length causes the $d_{x^2-y^2}$ orbital to shift down in energy, while the concomitant decrease in the tip--Fe distance increases the energy of the $d_{z^2}$ orbital [sketched in Figure~\ref{fig:dz}(c)].
The larger energy difference between $d_{\pi}$ and $d_{z^2}$ enhances also the energy difference $\text{E}_{1}-\text{E}_0$. This reduces the admixture of the excited states and, hence, the magnetocrystalline anisotropy. 
This scenario thus explains the initial slight increase in $D$ between  $\Delta z = 0$ and $-200\,{\rm pm}$ (Fe--surface distance increased and, therefore, lower $d_{z^2}$ energy)  and its subsequent pronounced decrease upon further tip approach.

\begin{figure}[t]
  \includegraphics[width=0.65\textwidth,clip=]{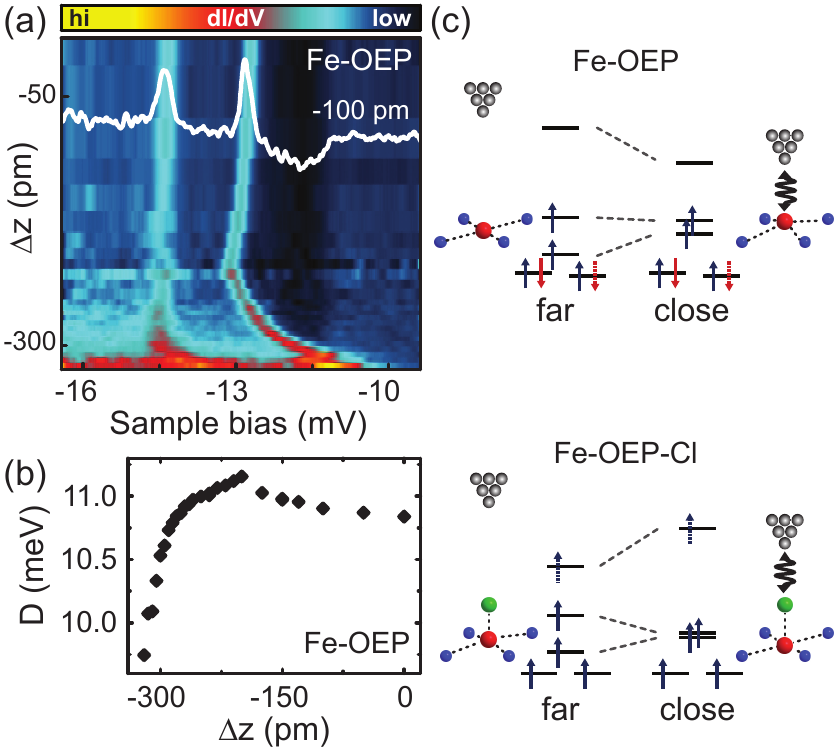}
  \caption{Tip-induced changes of the zero field splitting of Fe-OEP.
(a) Color plot of the normalized $dI/dV$ excitation spectra of Fe-OEP as a function of  $\Delta z$ ($50\,\mu {\rm V_{rms}}$, setpoint: $200\,{\rm pA}$, $50\,{\rm mV}$).
(b) Axial anisotropy $D$ \textit{vs.} $\Delta z$ as extracted from the spectra shown in (a). 
(c) Scheme of the changes in the ligand field splitting of Fe-OEP and FeOEP-Cl with the approaching tip.
}
\label{fig:dz}
\end{figure}

In the case of Fe-OEP-Cl, the tip potential acts (mainly) on the Cl ligand pointing upwards and attracts the Cl atom toward the tip. In turn, the Cl--Fe bond weakens and the Fe ion relaxes toward the molecular plane, decreasing the Fe--N bond length. The resulting changes of the ligand field splitting are sketched in Figure~\ref{fig:dz}(c)}: the $d_{z^2}$ ($d_{x^2-y^2}$) orbital shifts down (up) in energy as the Fe--Cl (Fe--N) distance increases (decreases). These changes increase the overall ligand field splitting and reduce the total energy difference $\text{E}_{1}-\rm{E}_0$, which also includes the spin pairing energy.
The smaller energy difference between ground and excited states enhances their admixture according to perturbation theory. This yields larger orbital moments $\Lambda_{xx}$ and $\Lambda_{yy}$ and, therefore, larger magnetocrystalline anisotropy.

The local control of the magnetic properties of nanostructures, such as the spin state or the magnetocrystalline anisotropy, is a prerequisite for their successful application in spintronic devices.  
In thin metallic films~\cite{weisheit07,maruyama09} and clusters~\cite{sonntag14}, an external electric field can be applied to tune the magnetic anisotropy. For single atoms, so far only static control of the magnetocrystalline anisotropy has been achieved by the selection of the adsorption site on surfaces~\cite{gambardella03,hirjibehedin07,loth10,rau14,bryant13}.  In the chemical approach, organic ligands are used to determine the ligand field splitting and magnetocrystalline anisotropy~\cite{gatteschi06}. Our approach provides the flexibility to continuously tune the magnetocrystalline anisotropy by modifying the geometry of the atomic-scale surrounding, even though it is, so far, restricted to changes in the order of 10\%. The opposed variations in the anisotropy induced by the tip for the two types of molecules discussed, underline the importance of a smart chemical engineering to achieve desired functionality and highlight the degree of freedom that can thereby be reached.
Furthermore, we show how the anisotropy may serve as a highly sensitive probe to identify variations in the atomic scale interactions, which are, {\it e.g.}, induced by the presence of the tip. 


\acknowledgement
We thank O. Peters and N. Hatter for assistance during magnetic-field dependent measurements. This research has been supported by the DFG grant FR2726/4, ERC grant NanoSpin, MINECO grant MAT2013-46593-C6-01, and the focus area "Nanoscale" of Freie Universit\"at Berlin.

\paragraph{Supporting Information Available}
Magnetic field dependent $dI/dV$ measurements on Fe-OEP-Cl.
This material is available free of charge via the Internet at http://pubs.acs.org.


\newpage
\section{Supporting Information:\\ Zeeman shift of the spin excitations}

\begin{figure}[hb]
  \includegraphics[width=1\textwidth,clip=]{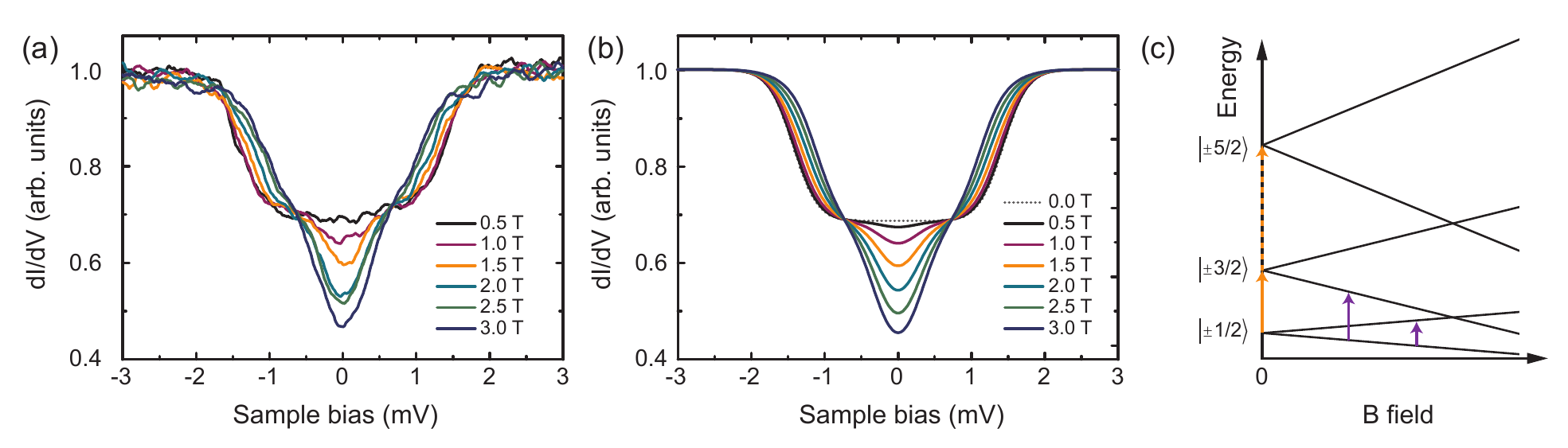}
  \caption{(a) Magnetic field dependent ISTS spectra of Fe-OEP-Cl (feedback parameters: $I=50$~mV, $V=200$~pA; $dz=-100$~pm, $V_{mod}=35~\mu$eV). (b) Fit of the experimental data in (a) with the following parameters: spin $S=5/2$, axial anisotropy $D=0.72\pm0.02$~meV,  g-factor $g=1.8\pm0.2$, effective temperature $T_{eff}=1.3$~K. (c) Scheme of the B field dependent state energies. Inelastic transitions as observed in the experiment are indicated by arrows.
}
\label{fig:Bfield}
\end{figure}

To prove the magnetic origin of the inelastic excitation detected for Fe-OEP-Cl, we acquired $dI/dV$ spectra at different magnetic (B) field strengths aligned perpendicular to the sample surface. Figure~\ref{fig:Bfield}(a) shows $dI/dV$ spectra measured on an Fe-OEP-Cl molecule at fields ranging from $0.5$ to $3$~T. The B field quenches the superconducting state of sample and tip. Therefore, the inelastic excitations now appear as steps rather then as resonances at a threshold energy $|eV|=\varepsilon$ in the $dI/dV$ spectra.

 With increasing field strength, the excitation with a zero-field energy of $1.4$~meV as detected in the superconducting state [compare to Figure~2(a) of the main manuscript], moves to lower energies. Simultaneously, a V-shaped gap is opened around zero bias. The energy of this new excitation increases with the B field strength. The appearance as a dip is due to the overlap of steps at opposite energies, which are  Fermi-Dirac broadened at $1.2$~K. 
We can describe the spin excitations in an anisotropic environment and magnetic field by the following phenomenological Spin--Hamiltonian~\cite{gatteschi06}: 
\begin{equation}
\mathcal{H} = g\mu_B \vec{B} \cdot \vec{S} + DS^2_z + E(S^2_x-S^2_y).
\end{equation}
The first term yields the Zeeman splitting, with $g$ being the Land\'e g-factor, $\mu_B$ the Bohr magneton, $\vec{B}$ the magnetic field vector and $\vec{S}= (S_x, S_y, S_z)$ the spin operator. $D$ and $E$ are the axial and transverse anisotropy parameters. 
Setting the spin to $S=5/2$ and $E=0$~\cite{heinrich13NP}, we can simulate the experimental spectra assuming the axial anisotropy parallel to the applied B-field axis and an effective temperature $T_{eff}=1.3$~K. The fit yields $D=0.72\pm0.02$~meV and  $g=1.8\pm0.2$ and the simulated curves are shown in Figure~\ref{fig:Bfield}(b).

Figure~\ref{fig:Bfield}(c) presents a scheme of the B field-dependent state energies and the observed transitions. In the superconducting state of the substrate at zero field strength, two excitations can be observed: one ground state excitation (full orange arrow)  from $M_S=\ket{\pm 1/2}$ to $\ket{\pm 3/2}$, and, at higher currents, a second excitation from the first excited state $\ket{\pm 3/2}$ to $\ket{\pm 5/2}$ (dashed orange arrow).   
In magnetic fields above the critical field of Pb ($80$~mT), only ground state excitations are observed. The violet arrows indicate the transitions from $\ket{-1/2}$ to $\ket{+1/2}$, and from $\ket{-1/2}$ to $\ket{-3/2}$, respectively.

In the case of Fe-OEP, we were not able to detect unambiguous evidence of a Zeeman shift of the spin excitations in fields of up to $3$~T at a temperature of $1.2$~K. This is understood by taking the relevant energies of the system into account. 
In a $S=1$ system  in zero magnetic field, the transverse anisotropy $E$ mixes the pure spin states $\ket{+1}$ and $\ket{-1}$.
 An applied magnetic field in $z$ direction will reduce the mixing and restore the pure spin state if $g\mu_BB_z \gg E$.
At intermediate fields, the energy of the two states is given as $E_\pm=D/3\pm \sqrt{E^2+(g\mu_BB_z)^2}$. 
With a transverse anisotropy of $E=0.7$~meV (approximate median of all measured molecules), this results in changes of the excitation energies of $60~\mu$eV at $3$~T, which is small compared to the temperature broadening at $1.2$~K ($3.5k_BT=360~\mu$eV). 
Yet, the agreement of the deduced axial anisotropy with bulk measurements of other Fe$^{2+}$ porphyrins~\cite{boyd79,barrac70}, together with the absence of any excitation in this energy range in the spectrum of Fe-OEP-Cl, renders a vibrational origin of the observed inelastic excitations implausible and corroborates our interpretation as spin excitations of a $S=1$ system. 


\begin{thebibliography}{nnnyys}
\bibitem{heinrich13NP} Heinrich, B. W.; Braun, L.; Pascual, J. I.; Franke, K. J. {\it Nature Phys.} {\bf 2013,} {\it9,} 765.

\bibitem{gatteschi06}  Gatteschi, D.; Sessoli, R.; Villain, J. {\it Molecular Nanomagnets}, Oxford Univ. Press: Oxford, 2006.

\bibitem{affronte08} Affronte, M. {\it J. Mater. Chem.} {\bf 2009,} {\it19,} 1731.

\bibitem{gambardella03} Gambardella, P.; Rusponi, S.; Veronese, M.; Dhesi, S. S.; Grazioli, C.; Dallmeyer, A.; Cabria, I.; Zeller, R.; Dederichs, P.H.; Kern, K.; Carbone, C.; and Brune, H. {\it Science} {\bf2003,} {\it300,} 1130.

\bibitem{bryant13} Bryant, B.; Spinelli, A.; Wagenaar, J.\ J.\ T.; Gerrits, M.; and Otte, A.\ F. {\it Phys. Rev. Lett.} \textbf{2013,} \textit{111,} 127203.

\bibitem{miyamachi13} Miyamachi,  T.; Schuh,	T.; M\"arkl, T.;	Bresch, C.; Balashov, T.; St\"ohr, A.; Karlewski,	 C.; Andr\'e, S.; Marthaler, M.; Hoffmann, M.; Geilhufe, M.; Ostanin, S.; Hergert,	W.; Mertig, I.; Sch\"on, F.; Ernst, A.; Wulfhekel, W. {\it Nature} {\bf2013,} {\it 503,} 242. 

\bibitem{rau14} Rau, I.; Baumann, S.; Rusponi, S.; Donati, F.; Stepanow, S.; Gragnaniello, L.; Dreiser, J.; Piamonteze, C.; Nolting, F.; Gangopadhyay, S.; Albertini, O. R.; Macfarlane, R. M.; Lutz, C. P.; Jones, B. A.; Gambardella, P.; Heinrich, A. J.; Brune, H. {\it Science} {\bf2014,} {\bf344,} 988. 

\bibitem{wang93} Wang, D.--S.; Wu, R.; Freeman, A.\ J. {\it Phys. Rev. B} {\bf 1993,} {\it47,} 14932.

\bibitem{honolka12} Honolka, J.; Khajetoorians, A. A.; Sessi, V.; Wehling, T. O.; Stepanow, S.; Mi, J.-L.; Iversen, B. B.; Schlenk, T.; Wiebe, J.; Brookes, N. B.; Lichtenstein, A. I.; Hofmann, Ph.; Kern, K.; and  Wiesendanger,R.{\it Phys. Rev. Lett.} \textbf{2012,} \textit{108,} 256811.

\bibitem{gambardella09} Gambardella, P.; Stepanow, S.; Dmitriev, A.; Honolka, J.; de Groot, F. M. F.; Lingenfelder, M.; Gupta, S. S.; Sarma, D. D.; Bencok, P.; Stanescu, S.; Clair, S.; Pons, S.; Lin, N.; Seitsonen, A. P.; Brune, H.; Barth, J. V.; Kern, K. {\it Nature Mat.} {\bf2009,} {\it 8,} 189.  %

\bibitem{tsukahara09} Tsukahara, N.; Noto, K.; Ohara, M.; Shiraki, S.; Takagi, N.; Takata, Y.; Miyawaki, J.; Taguchi, M.; Chainani, A.; Shin, S.; Kawai, M. {\it Phys. Rev. Lett.,} \textbf{2009,} {\it102,} 167203. %

\bibitem{heinrich13NL}   Heinrich, B. W.; Ahmadi, G.; M\"uller, V. L.; Braun, L.; Pascual, J. I.; Franke, K. J. {\it Nano Lett.} \textbf{2013,} {\it13,} 4840.


\bibitem{bhandary11} Bhandary, S.; Ghosh, S.; Herper, H.; Wende, H.; Eriksson, O.; Sanyal, B. P. {\it Phys. Rev.  Lett.} {\bf 2011,} {\it 107,} 257202.

\bibitem{parks10} Parks, J. J.; Champagne, A. R.; Costi,  T. A.; Shum, W. W.; Pasupathy, A. N.; Neuscamman, E.; Flores-Torres, S.; Cornaglia, P. S.; Aligia, A. A.; Balseiro, C. A.; Chan, G. K.-L.; Abru\~{n}a, H. D.; Ralph, D. C. {\it Science} {\bf 2010,} {\it 328}, 1370.

\bibitem{limot03} Limot, L.; Maroutian, T.; Johansson, P.; and Berndt, R. {\it Phys. Rev. Lett.} {\bf 2003,} {\it91,} 196801.

\bibitem{neel08} N\'eel, N.; Kr\"oger, J.; Limot, L.; and Berndt, R. {\it Nano Lett.} {\bf2008,} {\it8,} 1291.

\bibitem{choi12} Choi, D.-J.; Rastei,  M. V.; Simon, P.; and Limot, L. {\it Phys. Rev. Lett.} {\bf2012,} {\it108,} 266803.

\bibitem{bork11} Bork, J.; Zhang, Y.-h.; Diekh\"oner, L.; Borda, L.; Simon, P.; Kroha, J.; Wahl, P.; and Kern, K. {\it Nature Phys.} \textbf{2011,} {\it7,} 901. 

\bibitem{yan14} Yan,	 S.; Choi, D.-J.; Burgess, J. A. J.; Rolf-Pissarczyk, S.; Loth, S. {\it Nature Nanotech.} {\bf2014,} {\it10,} 40.


\bibitem{note1} Pb is a type I superconductor with a critical temperature of $7.2\,{\rm K}$. At $1.2\,{\rm K}$ its energy gap has a width of $2\Delta = 2.7\,{\rm meV}$.



\bibitem{gopakumar12} Gopakumar, T. G.; Tang, H.; Morillo, J.; and Berndt, R. {\it J. Am. Chem. Soc.} \textbf{2012,} {\it134,} 11844. 

\bibitem{stroz12} Str\'o\.zecka, A.; Soriano, M.; Pascual, J. I.; and Palacios, J. J. {\it Phys. Rev. Lett.} \textbf{2012,} {\it109,} 147202.

\bibitem{note3} {The doubling of the size of the superconducting gap is due to the convolution of the BCS-like (Bardeen--Cooper--Schrieffer) density of states of tip and sample.
The shoulder-like features visible at $\pm 6.7$  and $\pm 11.2\,{\rm meV}$ are, again, also present in the spectrum of pristine Pb(111). These are signatures of the strong electron-phonon coupling in Pb and correspond to the transverse and longitudinal phonon bands, respectively~\cite{rowell63}.}

\bibitem{rowell63} Rowell, J. M.; Anderson, P. W.; and Thomas, D. E. {\it Phys. Rev. Lett.} \textbf{1963,} {\it10,} 334.
\bibitem{shiba68} Shiba, H. {\it Prog. Theor. Phys.} \textbf{1968,} {\it40,} 435.

\bibitem{franke11} Franke, K. J.; Schulze, G.; and Pascual, J. I. {\it Science} \textbf{2011,}{\it332,} 940.


\bibitem{nishio01} Nishio, T.; Yokoyama, S.; Sato, K.; Shiomi, D.; Ichimura, A. S.; Lin, W. C.; Dolphin, D.; McDowell, C. A.; and Takui, T. 
{\it Synthetic Metals} \textbf{2001,} {\it121,} 1820.
\bibitem{note2} The absence of any resonances in this energy range in the spectrum of Fe-OEP-Cl renders a vibrational origin implausible.

\bibitem{handbook} Kadish, K. M.; and Ruoff, R. S. in {\it The Porphyrin Handbook}; Kadish, K. M., Smith, K. M., Guilard, R., Eds.; Academic Press: San Diego, 2000; Vol. 3.

\bibitem{wende07} Wende, H; Bernien, M.; Luo, J.; Sorg, C.; Ponpandian, N.; Kurde, J.; Miguel, J.; Piantek, M.;  Xu, X.; Eckhold, Ph.; Kuch, W.; Baberschke, K.; Panchmatia, P. M.; Sanyal, B.; Oppeneer, P. M.; \& Eriksson, O. {\it Nature Mater.} \textbf{2007,} {\it6,} 516. 


\bibitem{bernien07} Bernien, M.; Xu, X.; Miguel, J.; Piantek, M.; Eckhold, Ph.; Luo, J.; Kurde, J.; Kuch, W.; Baberschke, K.; Wende, H.; and Srivastava, P. {\it Phys. Rev. B} \textbf{2007,} {\it76,} 214406.

\bibitem{barrac70} Barraclough, C. G.; Martin, R. L.; Mitra, S.; and Sherwood, R. C. {\it J. Chem. Phys.} \textbf{1970,} {\it53,} 1643. 

\bibitem{boyd79} Boyd, P. D. W.; Buckingham, D. A.; McMeeking, R. F.; and Mitra, S. {\it Inorg. Chem.} \textbf{1979,} {\it18,} 3585. 

\bibitem{huheey03} Huheey, J.; Keiter, E.; Keiter, R. {\it Anorganische Chemie}, 3rd. ed.; Walter de Gruyter: New York, 2003.


\bibitem{note9} Note that the ground state electronic configuration of Fe(II) porphyrins is under debate for decades. However, it is agreed that the $d_{x^2-y^2}$ orbital is high up in energy, while all other $d$ levels are close in energy and the actual alignment depends on subtle details~\cite{liao04}. The resulting $S=1$ and in-plane anisotropy are robust. 
Despite being simplified, our sketch of the electronic configuration allows an intuitive insight and is sufficient to explain our experimental observations.  
Note that also the other low-lying excitation ($d_\pi \rightarrow d_{xy}$) contributes to an in-plane orbital moment.

\bibitem{liao04} Liao, M.-S.; Kar, T.; Gorun, S. M.; and Scheiner, S. {\it Inorg. Chem.} {\bf 2004,} {\it43,} 7151.  

\bibitem{dai08} Dai, D.; Xiang, H.; and Whangbo, M. {\it J. Comput. Chem.} {\bf2008,} {\it29,} 2187.

\bibitem{note8} We repeated this experiment on six Fe-OEP molecules with different $D$ and $E$ parameters. We observe qualitatively the same $D$ {\it vs.} $\Delta z$ dependence. The  $E$ {\it vs.} $\Delta z$ dependence differs, however, qualitatively for different molecules and tips. $E$ is a measure of distortions in the system and is, therefore, expected to strongly depend on details in the adsorption geometry. $D$ depends less on these distortions.  

\bibitem{flechtner07} Flechtner, K.; Kretschmann, A.; Steinr\"uck, H.-P.; and Gottfried, J. M. \textit{J. Am. Chem. Soc.} \textbf{2007,} {\it 129,} 12110.

\bibitem{weisheit07} Weisheit, M.; F\"ahler,  S.; Marty, A.; Souche, Y.; Poinsignon, C.; Givord, D. {\it Science} {\bf2007,} {\it315,} 349.



\bibitem{maruyama09} Maruyama, T.; Shiota, Y.; Nozaki,  T.; Ohta, K.; Toda, N.; Mizuguchi, M.; Tulapurkar, A. A.; Shinjo, T.; Shiraishi, M.; Mizukami, S.; Ando, Y.; Suzuki, Y. \textit{Nature Nanotech.} \textbf{2009,} {\it4,} 158. 

\bibitem{sonntag14} Sonntag, A.; Hermenau, J.; Schlenhoff, A.; Friedlein, J. Krause, S.; and Wiesendanger, R. {\it Phys. Rev. Lett.} {\bf2014,} {\it112,} 017204.

\bibitem{hirjibehedin07} Hirjibehedin, C. F.; Lin, C.-Y.; Otte, A. F.; Ternes, M.; Lutz, C. P.; Jones, B. A.; and Heinrich, A. J. {\it Science} \textbf{2007,} {\it317}, 1199.

\bibitem{loth10} Loth, S.; Etzkorn, M.; Lutz, C. P.; Eigler, D. M.; Heinrich, A. J. {\it Science} \textbf{2010,} {\it329,} 1628.





\end{thebibliography}
\end{document}